\documentclass[conference]{IEEEtran}
\IEEEoverridecommandlockouts
\usepackage{cite}
\usepackage{amsmath,amssymb,amsfonts}
\usepackage{algorithmic}
\usepackage{graphicx}
\usepackage{textcomp}
\usepackage{tabularx}
\usepackage{url}
\usepackage{xcolor}
\def\BibTeX{{\rm B\kern-.05em{\sc i\kern-.025em b}\kern-.08em
    T\kern-.1667em\lower.7ex\hbox{E}\kern-.125emX}}
\begin{document}

\title{Embedded Firmware Development for Flight Control, Telemetry, and Video Streaming for DIY UAV Research\\
}

\author{
\begin{tabular}{@{}p{0.3\textwidth} @{\hskip 0.05\textwidth} p{0.3\textwidth} @{\hskip 0.05\textwidth} p{0.35\textwidth}@{}}
\centering
\textbf{1\textsuperscript{st} Ad-Deen Mahbub} \\
Dept. of Electrical \& Electronic Engineering \\
Shahjalal University of Science \& Technology \\
Sylhet 3100, Bangladesh \\
addeenmahbub@gmail.com
&
\centering
\textbf{2\textsuperscript{nd} Indroneel Roy} \\
Dept. of Electrical \& Electronic Engineering \\
Shahjalal University of Science \& Technology \\
Sylhet 3100, Bangladesh \\
indroneel@example.com
&
\centering
\textbf{3\textsuperscript{rd} Dr. Mohammad Kamruzzaman Khan Prince} \\
Associate Professor \\
Dept. of Electrical \& Electronic Engineering \\
Shahjalal University of Science \& Technology \\
Sylhet 3100, Bangladesh \\
kamruzzaman-eee@sust.edu
\end{tabular}
}

\maketitle

\begin{abstract}
This paper presents the design and optimization of quadcoptor firmware for a low-cost UAV based on the dual-core ESP32-S3 microcontroller, targeting real-time flight control, telemetry, and video transmission within stringent embedded resource constraints. The firmware leverages dedicated task allocation across the two cores: 400Hz control loop on Core 1 implements sensor fusion with a Madgwick filter, a hybrid fixed-point PID controller enhanced by selective derivative scaling and feedforward braking. Core 2 handles 50 Hz telemetry encoding and UDP-based video and data streaming at 25FPS VGA frames, ensuring separation of control and communication workloads to prevent latency interference. Inter-core communication employs zero-copy buffers with mutex protection for minimal overhead and deterministic execution. Custom event-driven threading and RTOS-inspired static scheduling guarantee real-time responsiveness and jitter-free operation despite ESP32-S3’s limited memory and floating-point performance. Experimental evaluation confirms that this firmware architecture provides precise, stable flight stabilization and robust telemetry support, enabling resource-constrained ESP32-S3 platforms to handle DIY projects and research with minimal investment and full customization capability.
\end{abstract}

\begin{IEEEkeywords}
ESP32-S3 Flight Control Firmware, Telemetry Transmission, Video Streaming, Sensor Fusion, Hybrid Cascaded PID Control with Feedforward Braking, Real-Time Operating System (RTOS).
\end{IEEEkeywords}

\section{Introduction}
Unmanned aerial vehicles (UAVs) are rapidly transforming numerous sectors, including environmental monitoring, disaster response, agriculture, logistics, and surveillance, making UAV technology critically important for the coming decades. Despite this growing significance, widespread research and innovation in UAVs face substantial financial barriers \cite{b1,b2}. Existing UAV systems often rely on expensive sensors, high-performance embedded processors, and commercial firmware solutions that come with restrictive licensing and limited access to low-level customization\cite{b3,b4}. These factors significantly limit the adaptability and extendability of UAV platforms for researchers, educators, and hobbyists operating under constrained budgets\cite{b5,b6}.

Market-ready UAV firmwares typically offer flexibility for modification at the core control and telemetry layers,but most of them are built for arm - cortex architecture which do not provide built in wifi communication channel like Esp32S3 does and rely on external telemetry and Vtx modules increasing payload, energy consumption and closed modularity in data transmission system, hindering the ability of DIY researchers and small-scale developers to tailor system behavior or integrate novel algorithms such as cutting-edge simultaneous localization and mapping (SLAM) methods \cite{b7,b8}. This lack of firmware transparency and customizability creates a bottleneck that slows down innovation and limits educational outreach.

To bridge this gap, it is essential to democratize UAV technology by providing open, low-cost, and fully customizable platforms that empower researchers and students worldwide \cite{b9}. There are exisitng firmwares that offer opensource firmware but lacks full customization capability and ease of access and requires expensive and energy consuming communication module interface. This work compliments the gap by developing and open-sourcing a UAV system based on a low-cost dual-core ESP32-S3 microcontroller, optimized scalable firmware, and an external PC-WiFi interface for computationally intensive task offloading for further advanced UAV research. This platform enables anyone with limited resources to build, program, and modify miniaturized  drones at the firmware level, eliminating reliance on costly, closed-access systems.

By releasing the firmware and design openly, it was an attempt to foster a collaborative ecosystem where users can adapt the control, telemetry, and video transmission functions according to their specific research needs, this approach lowers the entry barrier, accelerates innovation, and cultivates a new generation of UAV developers prepared to tackle diverse real-world challenges with accessible, scalable drone technology\cite{Patil2024,b,Kim2024,Glossner2021}.

\section{Methodology}

This section focuses on two critical components of the UAV firmware: \textit{Control Structure} and \textit{Telemetry Optimization and Video Streaming}. Each subsection is organized as key problems encountered and corresponding solutions implemented.

\subsection{Control Structure}

\paragraph{Problem 1: Shortcomings of Classical PID in Systems with Momentum \cite{Guzman2021}}  
Classical PID controllers often fail to stabilize UAVs effectively due to system momentum and nonlinearities affecting roll, pitch, and yaw dynamics. Single-loop PID may accumulate errors leading to sluggish and oscillatory flight behavior.

\paragraph{Solution: Cascading PID with Feedforward Braking}  
To address this, a cascading PID control structure with three independent PID controllers was designed, managing roll, pitch, and yaw separately. This separation enables tailored tuning for each axis. Additionally, a feedforward braking model anticipates rotational velocity and angular position, applying pre-emptive braking with a tunable parameter analogous to PID gains.

\begin{figure*}[ht]
    \centering
    \includegraphics[width=0.98\textwidth]{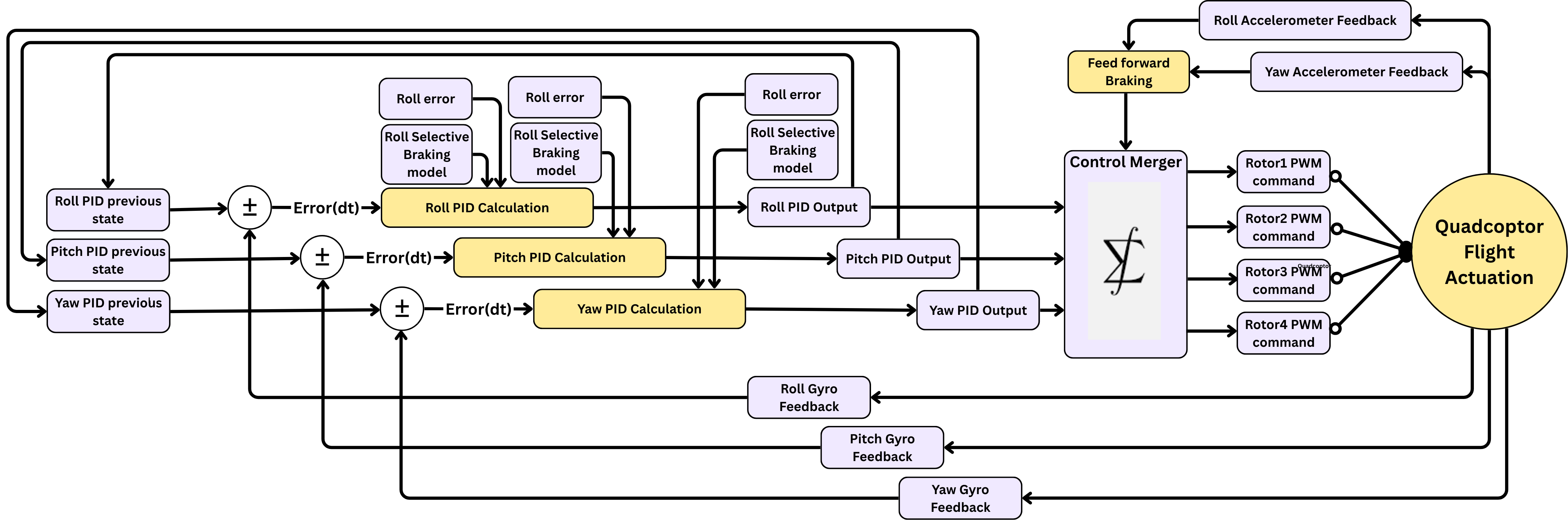}
    \caption{UAV Hybrid Control Loop with feed forward braking and selective derivative modeling}
    \label{UAV Hybrid Control}
\end{figure*}

A proactive braking mechanism demonstrated in Figure~\ref{UAV Hybrid Control} was introduced. Specifically, when the error derivative is negative and the absolute error is decreasing (i.e., the current state is approaching the target), a one-time braking pulse is applied proportional to the rate of convergence in \eqref{eq:braking}:

\begin{equation}
\label{eq:braking}
u_b = -k_b \cdot \left|\frac{de(t)}{dt}\right| \quad \text{if } e(t) > 0 \text{ and } \frac{de(t)}{dt} < 0
\end{equation}

This feedforward braking term \( u_b \) is applied only once per convergence event and acts as a temporary suppressive force, altering the system dynamics so that the remaining correction can be handled with lighter derivative influence. As a result, the proportional gain \( K_p \) can be increased more aggressively without inducing oscillations.

\paragraph{Problem 2: Sensor Jitter Causing High Derivative Kick Close to Zero Error \cite{Lendek2021,Kumar2016}}  
Sensor noise around the stable state can cause noisy derivative terms, known as derivative kick, destabilizing the control loop during steady flight.

\paragraph{Solution} Let \( \sigma \) denote the \textit{influence spread}, a tunable constant that controls the width of the modulation curve. In \eqref{eq:Sx} \& \eqref{eq:Sy} smaller values of \( \sigma \) increase the effective spread of derivative influence across a wider error range, while larger values make the modulation more selective, confining derivative influence to small-error regions.

\begin{align}
\label{eq:Sx}
S_x(e) ={}& \min\Big(1.0,\ \max\Big(0.0,\ \notag \\
        & \quad \frac{10}{\left(|e - 3| \cdot |e + 3| \cdot \sigma_x\right) + 1} \cdot 5 \Big)\Big)
\end{align}

Where \( \sigma_x = 0.2 \) is the influence spread for the roll axis.

Similarly, for pitch stabilization (Y-axis), the modulation factor \( S_y(e) \) was given by:

\begin{align}
\label{eq:Sy}
S_y(e) ={}& \min\Big(1.0,\ \max\Big(0.0,\ \notag \\
        & \quad \frac{10}{\left(|e - 5| \cdot |e + 5| \cdot \sigma_y\right) + 1} \cdot 5 \Big)\Big)
\end{align}

Where \( \sigma_y = 0.067 \) is the influence spread for the pitch axis.

\begin{figure}[ht]
    \centering    \includegraphics[width=0.48\textwidth]{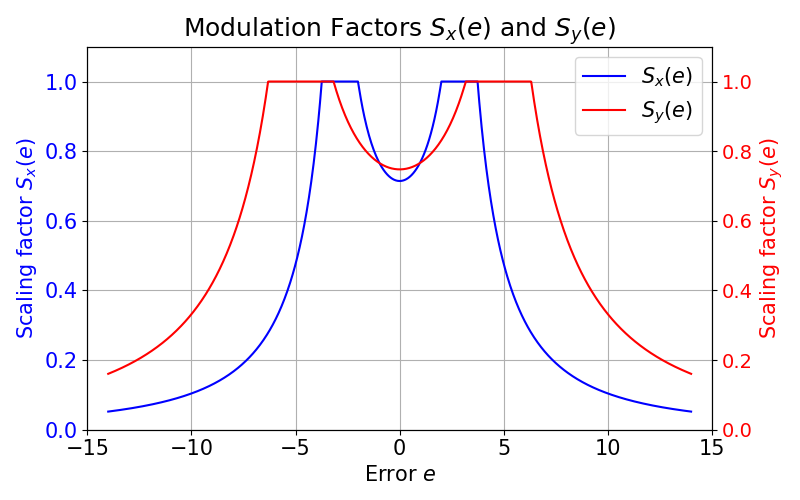}
    \caption{Selective Roll correction derivative scale
model}
    \label{fig:Selx}
\end{figure}

Where \( S_{axis}(e) \) is either \( S_x(e) \) or \( S_y(e) \) depending on the control axis. This formulation allowed aggressive use of the proportional term during large-error phases while ensuring that derivative damping becomes effective only when convergence is near, leading to smoother and faster stabilization demonstrated in Figure~\ref{fig:Selx}.

These functions are bounded between 0 and 1 using a clamping mechanism to prevent instability. The modulation factor peaks around \( e = \pm3 \) for roll and \( e = \pm5 \) for pitch — regions where precise damping is essential — and decays rapidly as the error magnitude increases. This approach suppresses derivative opposition during high-error states, allowing the proportional term to dominate, and enables fine-grained derivative correction near the target.

The final derivative term is then expressed as:

\begin{equation}
D(t) = K_d \cdot S_{axis}(e) \cdot \frac{de(t)}{dt}
\end{equation}

The hybrid control law produces individual control outputs for roll, pitch, and yaw axes as:

\begin{align}
u_i(t) ={}& K_{p,i} \cdot e_i(t) 
        + K_{i,i} \int e_i(t) \, dt \notag \\
        &+ K_{d,i} \cdot S_i(e_i(t)) \cdot \frac{de_i(t)}{dt} \notag \\
        &+ u_{b,i}(t), \quad i \in \{x, y, z\}
\end{align}
where \( u_i(t) \) is the control signal along axis \(i\); 
\( e_i(t) \) is the tracking error along axis \(i\); 
\( K_{p,i} \), \( K_{i,i} \), and \( K_{d,i} \) are the proportional, integral, and derivative gains, respectively; 
\( S_i(e_i(t)) \) is a nonlinear scaling function applied to the derivative term; 
\( u_{b,i}(t) \) is the feedforward or bias input; 
and \( i \in \{\text{roll}, \text{pitch}, \text{yaw}\} \) denotes the three rotational degrees of freedom of the system.

\paragraph{Problem 3: CPU Starvation Due to Busy Loops Without Yielding \cite{noauthor_undated}}  
Without calls to functions such as \texttt{delay()}, \texttt{yield()}, or \texttt{vTaskDelay()}, the CPU remains occupied executing tight loops, starving FreeRTOS background tasks responsible for WiFi, Bluetooth, and housekeeping, potentially causing device unresponsiveness.

\begin{figure*}[h]
    \centering
    \includegraphics[width=0.98\textwidth]{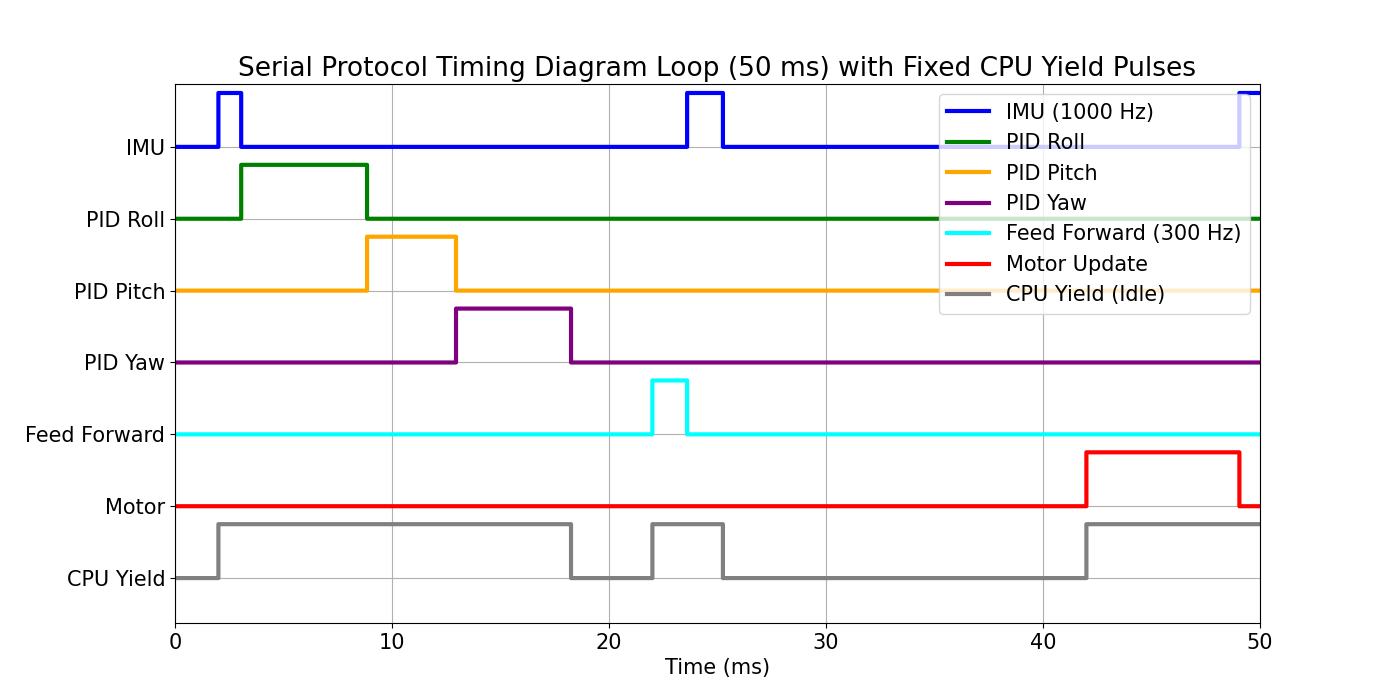}
    \caption{Timing Diagram of parallel threaded processes}
    \label{Timing Diagram of parallel threaded processes}
\end{figure*}

\paragraph{Solution: Adaptive Yielding Strategies}  
The firmware design incorporates adaptive yield calls within control loops running at different frequencies (sensor data fetching, motor signal updates) to allow FreeRTOS background tasks execution. This approach prevents watchdog timeouts and ensures system responsiveness demonstrated in Figure~\ref{Timing Diagram of parallel threaded processes}.

\subsection{Telemetry and Computation Optimization}

\paragraph{Problem 4: Control Loop and Communication Interference \cite{Santoso2018}}  
Executing WiFi telemetry communication within the same thread or core as the control loop can induce latency and jitter, destabilizing flight operation.

\begin{figure}[h]
    \centering
    \includegraphics[width=\linewidth]{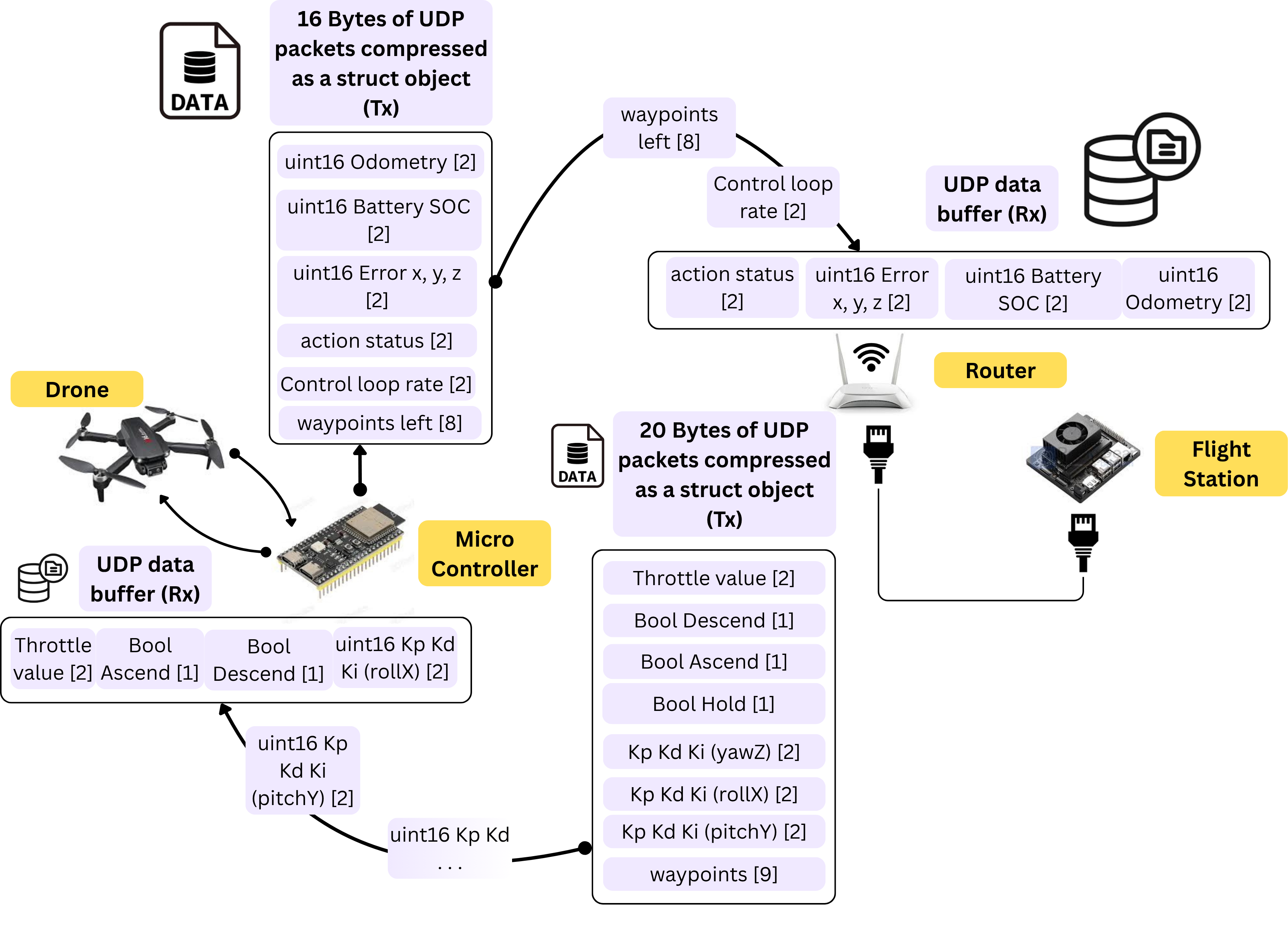}
    \caption{Communication architecture and UDP data flow between UAV and
ground station}
    \label{fig:UDP telemetry}
\end{figure}

\paragraph{Solution: Dual-Core Task Segregation}  
 ESP32-S3's dual-core architecture was leveraged by dedicating one core to a 400 Hz flight control loop and the other to telemetry encoding and UDP transmission at 50 Hz. This separation minimizes interference, yielding consistent control and communication performance. The overview of the system demonstrated in Figure~\ref{fig:UDP telemetry}.

\paragraph{Problem 5: UDP Packet Loss and Telemetry Corruption}  
UDP is unreliable due to possible packet loss, which can corrupt telemetry messages if spread across multiple packets.
\paragraph{Solution: Compact Single-Byte Telemetry Messages}  
Telemetry data is compacted into UDP packet, ensuring each received packet contains a complete telemetry snapshot. This eliminates the need for complex synchronization or retransmissions and gracefully tolerates dropped packets without partial data reception. A single udp packet contain upto 65,535 bytes of data which was more than enough to communicate with the drone, when each byte can be compacted to transmit on block of control command.

\subsection{VGA Video Streaming over UDP: Challenges and Solutions}

Streaming VGA resolution (640$\times$480) video at real-time frame rates over UDP presents inherent challenges due to the protocol’s unreliability and asynchronous nature. UDP does not guarantee packet delivery, order, or frame integrity, which causes issues like dropped packets, lost frames, and corrupted image reconstruction.

\paragraph{Key Problems:}
\begin{itemize}
    \item \textbf{Packet loss leading to complete frame drops:} UDP packets carry segments of JPEG-compressed frames; losing a single packet may invalidate the whole frame, resulting in visible glitches or freezes.
    \item \textbf{Asynchronous transmission and synchronization requirements:} Video packets arrive out-of-order and with varying latency, necessitating explicit synchronization to reconstruct frames correctly.
    \item \textbf{Frame misconstruction:} Incomplete or partial frames from missing packets can cause decoder errors or visual artifacts at the receiver.
\end{itemize}

\paragraph{Solution Pipeline}

\begin{figure}[h]
    \centering
    \includegraphics[width=\linewidth]{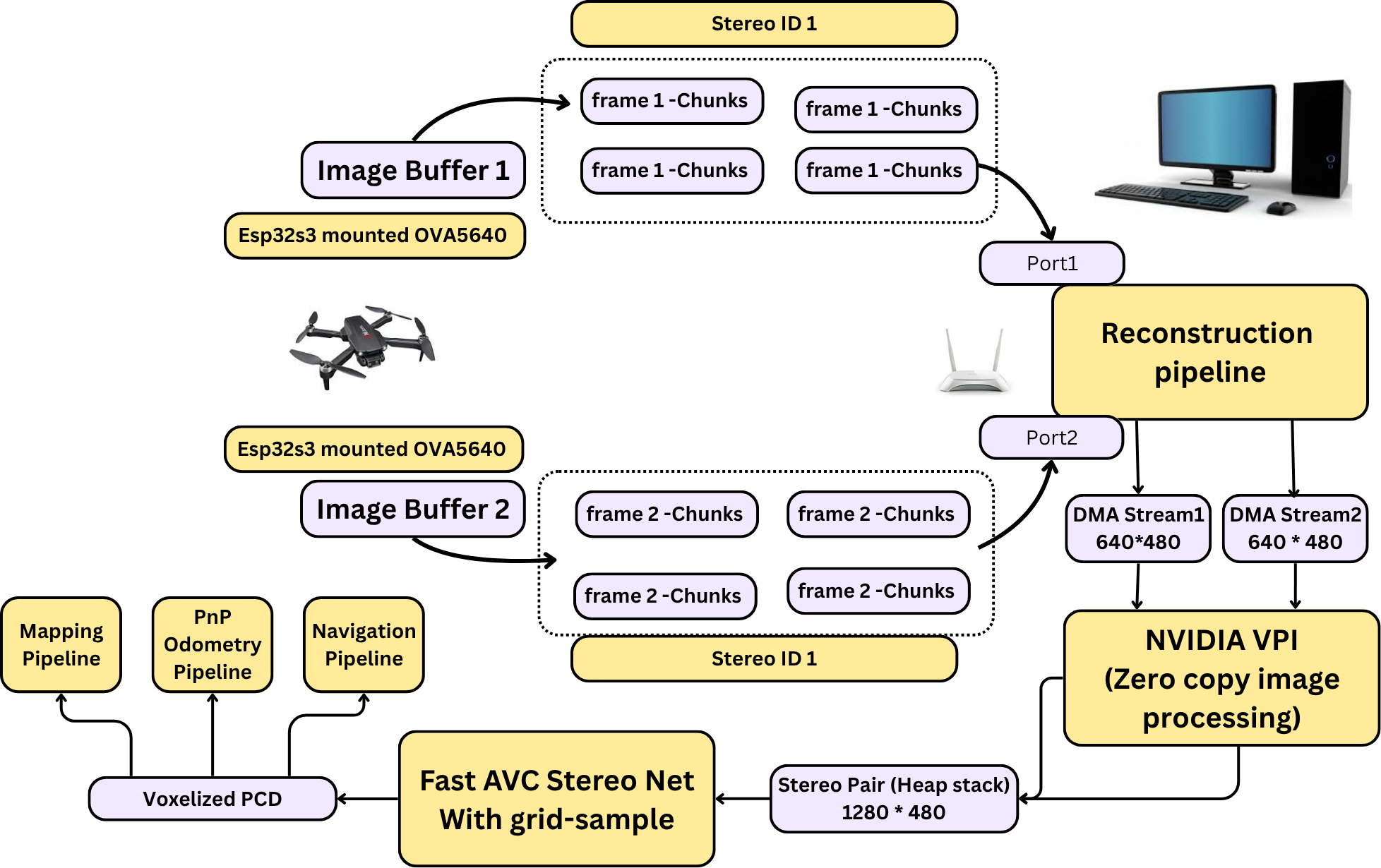}
    \caption{Video Streaming pipeline}
    \label{fig:Video Streaming pipeline}
\end{figure}

\textbf{Hardware and Firmware Integration:}  
OV5640 5 MP CMOS image sensor was utilized with built-in JPEG compression, interfaced to an ESP32-S3 CAM WROOM module featuring a dedicated camera interface (DCMI) and sufficient external PSRAM. Due to lack of native driver support on low-cost ESP32-S3 clone boards, configuration from OV5640 datasheets and existing ESP32-S3 camera libraries was manually merged. This enabled continuous VGA JPEG frame capture and buffering (frame sizes 8--25\,KB). Full overview is shocased in Figure~\ref{fig:Video Streaming pipeline}.

\textbf{Frame Buffering and UDP Segmentation:}  
Captured frames are stored in a PSRAM FIFO buffer and segmented into fixed-size UDP packets (1000 bytes each), resulting in roughly 10--25 packets per frame. Each packet includes a \textit{Frame ID} and \textit{Patch ID} for proper grouping and reordering at the receiver.

\textbf{Robust Frame Reassembly Under Packet Loss:}  
To mitigate dropped frames due to packet loss, a lightweight blur-padding strategy was adopted. Missing chunks are replaced with blank patches locally blurred via Gaussian filtering, preserving image coherence and minimizing decoding errors without complex interpolation~\cite{Katiyar2024}.

\textbf{Dual-Core Processing and Memory Optimization:}  
The ESP32-S3 dual-core architecture was exploited by assigning one core to high-speed image acquisition and PSRAM buffering with overflow control, and the other core to JPEG chunk preparation and UDP transmission. Marking buffers \texttt{volatile} ensured timely memory clearance, reducing lag and bounding memory usage~\cite{Ratti2011}.

This integrated hardware-software approach addresses UDP streaming challenges—packet loss, asynchronous arrival, and frame misconstruction—through a carefully designed pipeline.

\section{Experimental Setup}
\begin{figure}[h]
    \centering
    \includegraphics[width=1\linewidth]{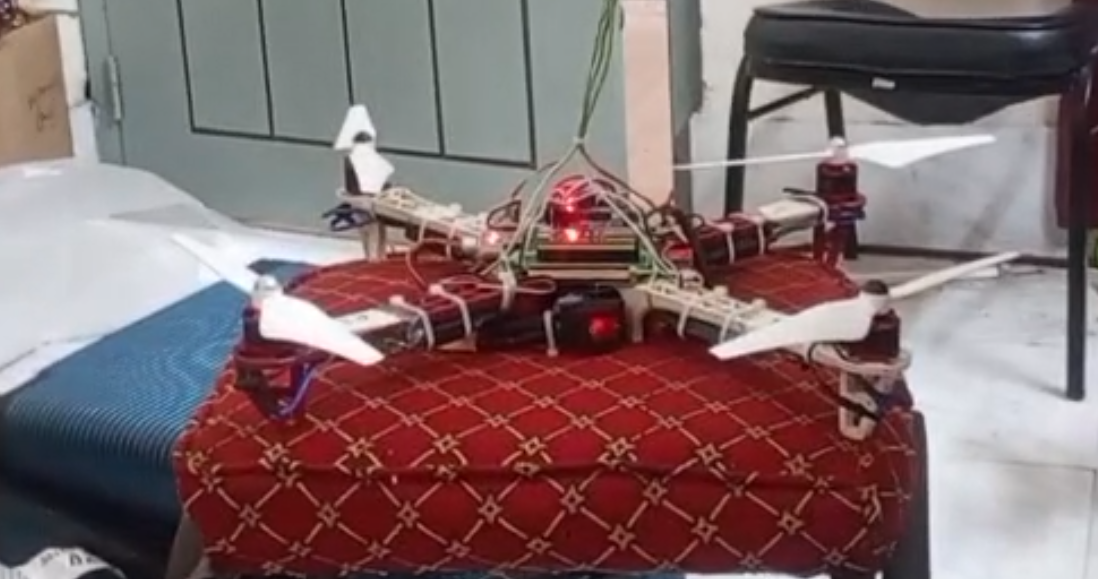}
    \caption{DIY Drone clamped from above}
    \label{fig:DIY Drone clamped from above}
\end{figure}

To evaluate the UAV firmware control performance, a custom low-cost DIY quadcopter platform based on the ESP32-S3 microcontroller and OV5640 camera module was contructed. To ensure controlled and reproducible testing conditions without risk of uncontrolled flight, the drone was securely clamped from above shown in  Figure~\ref{fig:DIY Drone clamped from above}.

\section{Results \& Discussions}

\begin{figure}[h]
    \centering
    \includegraphics[width=1\linewidth]{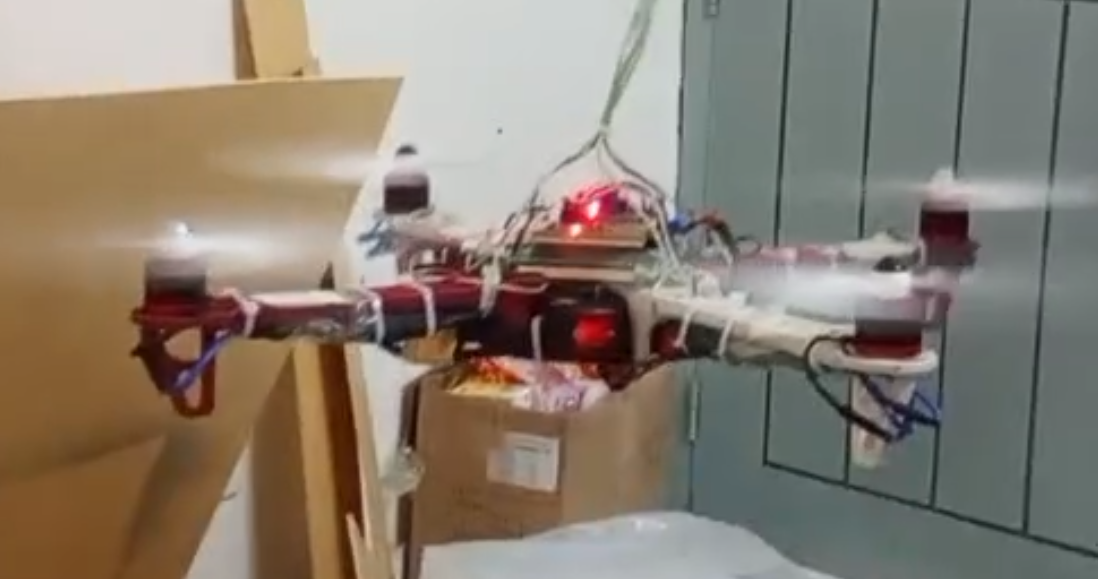}
    \caption{DIY Drone Hover state testing}
    \label{fig:DIY Drone Hover state testing}
\end{figure}

\begin{figure}[h]
    \centering
    \includegraphics[width=\linewidth]{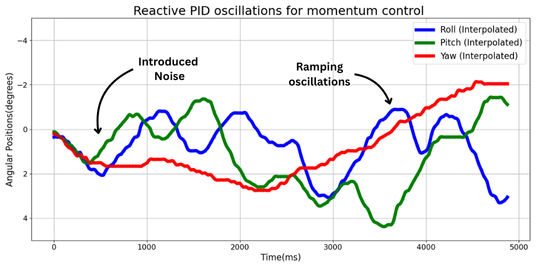}
    \caption{Reactive problem of controlling momentum with PID}
    \label{fig:Momentum Oscillations}
\end{figure}

\begin{figure}[h]
    \centering
    \includegraphics[width=\linewidth]{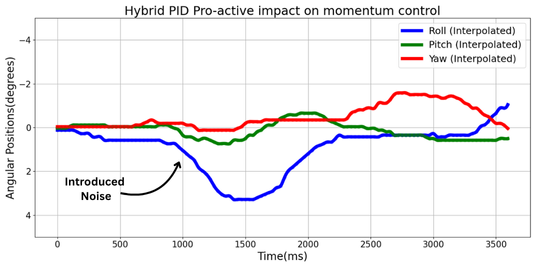}
    \caption{Proactive braking results of controlling momentum with hybrid PID with feedforward braking model}
    \label{fig:Proactive braking}
\end{figure}

A series of flight tests was conducted, shown in in Figure~\ref{fig:DIY Drone Hover state testing} by commanding flight inputs via telemetry from command station over wifi, comparing the classical reactive PID controller with the efficient hybrid controller integrating feedforward braking and selective nonlinear derivative scaling demonstrated in Figure~\ref{fig:Momentum Oscillations} \&  ~\ref{fig:Proactive braking}.

To simulate real-world disturbances and evaluate robustness, external noise was applied by manually agitating the drone stable hovering state with strings attached to the frame during neutral and commanded states.
With standard PID control the drone hovered successfully, but the problem arises when the drone is agitated mid flight (via strings) and PID struggles to stabilize momentum causing multiple overshoots \& mis-positioning the drone. To overcome these shortcomings, a feedforward braking model was integrated within the PID control framework, forming a hybrid control network demonstrated in Figure~\ref{UAV Hybrid Control}. 

This feedforward element predicted rotational velocity and current angular position to proactively apply braking forces before large deviations could occur. The effect was a significant improvement in stability and control responsiveness, reducing overshoot and oscillations notably in aggressive flight tests in ~\ref{fig:Proactive braking}. This tests demonstrated that this combined PID and feedforward control approach enabled sharper, more accurate maneuvering and faster convergence to setpoints compared with reactive PID control in Figure \ref{fig:Momentum Oscillations} alone, which was particularly evident in roll and pitch stabilization during rapid setpoint changes.

\begin{figure}[h]
    \centering
    \includegraphics[width=\linewidth]{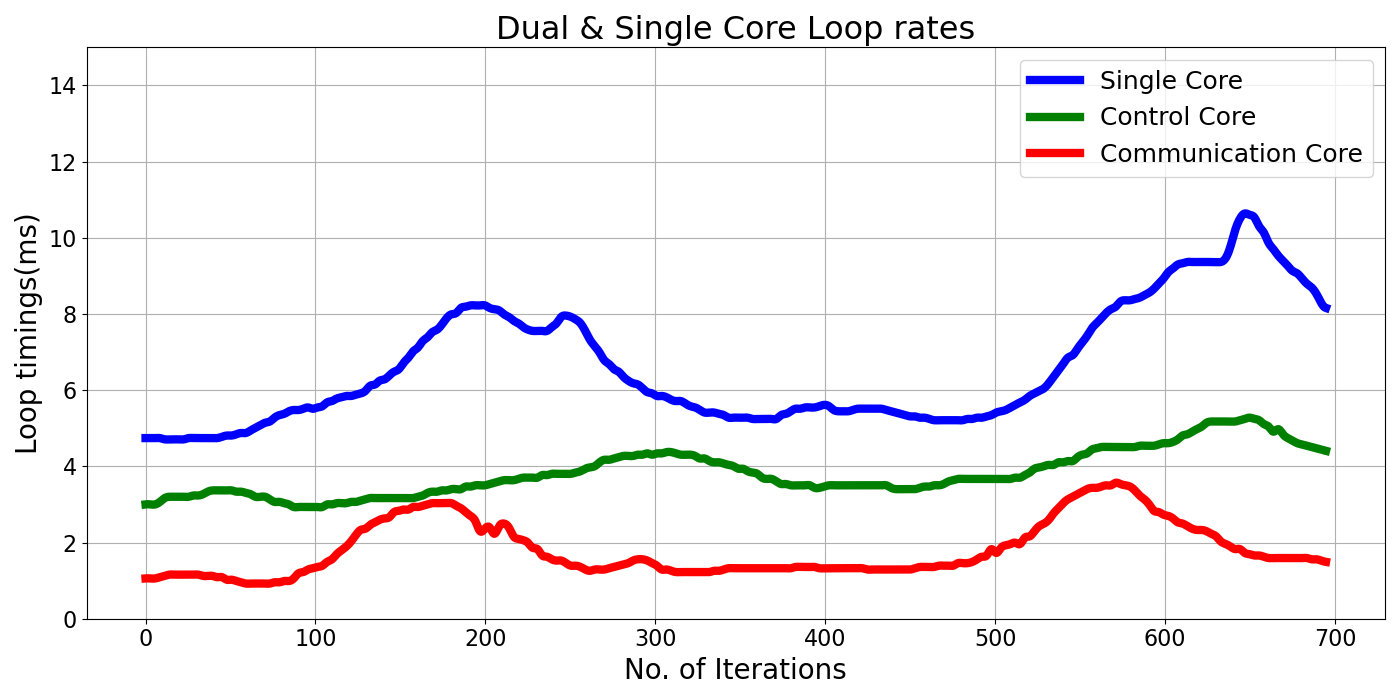}
    \caption{Loop rate shifts over iterations for Single Core And Dual core computational load distribution}
    \label{fig:Loop rate shifts}
\end{figure}

\begin{figure}[h]
    \centering
    \includegraphics[width=\linewidth]{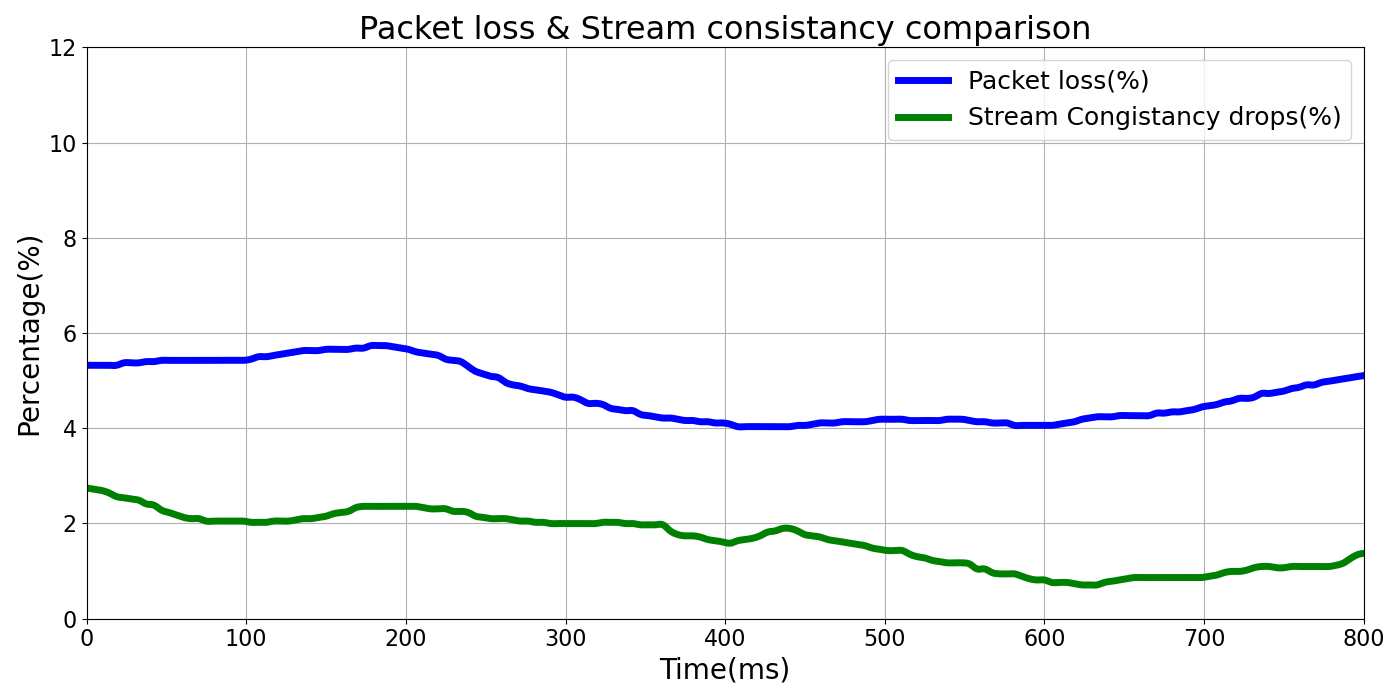}
    \caption{The robustness of stream consistancy not increasing with fluctuating packet loss(\%)}
    \label{fig:he robustness of stream}
\end{figure}

The ESP32-S3’s dual-core architecture separated tasks with dedicated timing. The control core ran sensor fusion, PID, and feedforward loops at 400 Hz for real-time flight control, while the second core handled telemetry and video streaming at 50 Hz, ensuring stable UDP transmission without disrupting control. This design maintained both control precision and reliable communication under the system's strict computational limits, as shown in Figure~\ref{fig:he robustness of stream}.

This combination of improved control law and precise timing hierarchy strongly validates the firmware design choices for low-cost, resource-constrained UAV autonomy using the ESP32-S3 microcontroller.

\begin{figure}
    \centering
    \includegraphics[width=\linewidth]{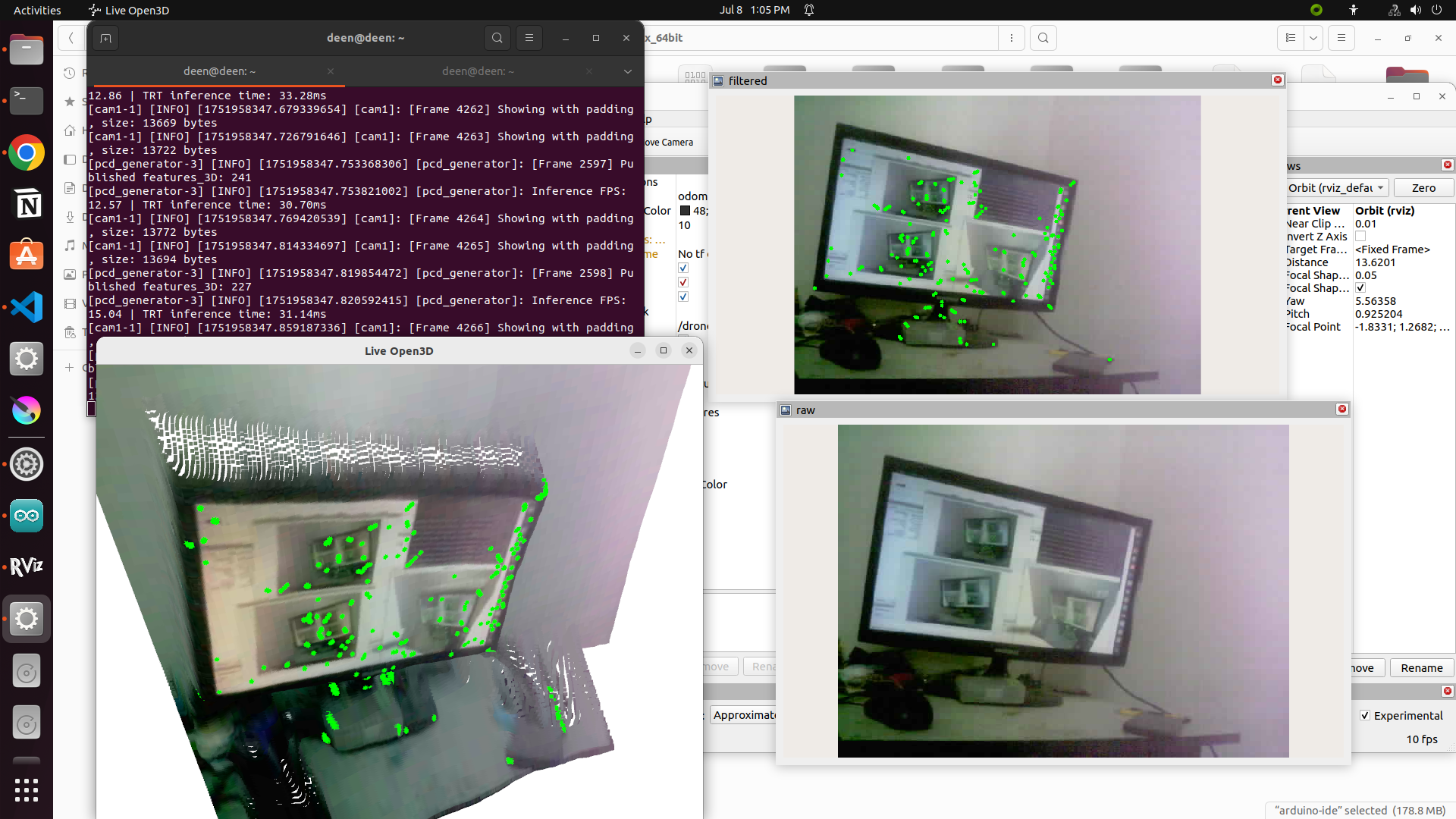}
    \caption{Live video transmission from the drone back to command station interfaced with ROS
architecture}
    \label{fig:Live video transmission from the drone back to command station interfaced with ROS
architecture}
\end{figure}

\section{Conclusion}
This drone firmware, developed on the ESP32 platform, significantly lowers the entry barrier compared to widely-used ARM Cortex-based open source firmwares such as ArduPilot and PX4. By leveraging the ESP32's integrated WiFi and camera modules, the approach consolidates flight control, telemetry, and video streaming within a single low-cost microcontroller, programmed using Arduino-compatible code with full source accessibility. This integration eliminates the need for external telemetry radios and video transmitters, which are typically required in ARM Cortex-based systems, thus reducing hardware complexity, cost, and power consumption. While ARM Cortex MCUs inherently provide robust real-time flight control and extensive peripheral support with industrial-grade reliability, they generally require additional specialized communication hardware and involve a moderate learning curve to navigate their sophisticated modular firmware architectures. In contrast, the ESP32’s dual-core Tensilica processor and rich connectivity suite enable optimized real-time performance when properly programmed, along with seamless wireless telemetry and live video streaming capabilities. As a result, the system facilitates rapid prototyping and cost-effective design, democratizing access to drone development without compromising essential functionality. This approach aligns with recent trends emphasizing integrated wireless communication in drone platforms \cite{FlywingTech2024, Wik2025}, presenting a compelling alternative to traditional multi-module setups.


\begin{thebibliography}{00}
\bibitem{b1}
C. Cadena, L. Carlone, H. Carrillo, Y. Latif, D. Scaramuzza, J. Neira, I. Reid, and J. J. Leonard, “Past, present, and future of simultaneous localization and mapping: Towards the robust-perception age,” \textit{IEEE Transactions on Robotics}, vol. 32, no. 6, pp. 1309–1332, Dec. 2016.

\bibitem{b2}
Walker, S. E.; Sheaves, M.; Waltham, N. J. ``Barriers to using UAVs in conservation and environmental management: A systematic review.'' \textit{Environmental Management}, vol. 71, no. 5, pp. 1052--1064, May 2023. Springer Science and Business Media LLC.

\bibitem{b3}
Kamat, A.; Shanker, S.; Barve, A.; Muduli, K.; Mangla, S. K.; Luthra, S.  
Uncovering interrelationships between barriers to unmanned aerial vehicles in humanitarian logistics.  
\textit{Operations Management Research}, vol. 15, no. 3-4, pp. 1134--1160, Dec. 2022.  
Springer Science and Business Media LLC.  
\url{https://www.springernature.com/gp/researchers/text-and-data-mining}

\bibitem{b4}
Singh, A. K.; Mohandes, S. R.; Muhodir, S. H.; Zhang, W.; Antwi-Afari, M. F.; Shakor, P.  
Exploring barriers to unmanned aerial vehicle (UAV) technology for construction safety management using mixed-methods approach.  
\textit{Buildings}, vol. 15, no. 12, Art. no. 2092, Jun. 2025.  
MDPI AG.  
\url{https://creativecommons.org/licenses/by/4.0/}

\bibitem{b5}
Holton, A. E.; Lawson, S.; Love, C.  
Unmanned Aerial Vehicles.  
\textit{Journal of Practice}, vol. 9, no. 5, pp. 634--650, Sep. 2015.  
Informa UK Limited.

\bibitem{b6}
Stöcker, C.; Bennett, R.; Nex, F.; Gerke, M.; Zevenbergen, J.  
Review of the current state of UAV regulations.  
\textit{Remote Sensing (Basel)}, vol. 9, no. 5, Art. no. 459, May 2017.  
MDPI AG.  
\url{https://creativecommons.org/licenses/by/4.0/}

\bibitem{b7}
Ozbiltekin-Pala, M.; Yavas, V.; Ozkan-Ozen, Y. D.  
Drivers and barriers of unmanned aerial vehicles in emergency logistics operations.  
\textit{Technology in Society}, vol. 82, Art. no. 102894, Sep. 2025.  
Elsevier BV.

\bibitem{b8}
Sørensen, L.; Jacobsen, L.; Hansen, J.  
Low cost and flexible UAV deployment of sensors.  
\textit{Sensors (Basel)}, vol. 17, no. 1, Art. no. 154, Jan. 2017.  
MDPI AG.  
\url{https://creativecommons.org/licenses/by/4.0/}

\bibitem{b9}
Ebeid, E.; Skriver, M.; Terkildsen, K. H.; Jensen, K.; Schultz, U. P.  
A survey of open-source UAV flight controllers and flight simulators.  
\textit{Microprocessors and Microsystems}, vol. 61, pp. 11--20, Sep. 2018.  
Elsevier BV.

\bibitem{FlywingTech2024}
Flywing Tech.  
STM32 vs ESP32: Which Microcontroller Is Right for Your Project? [Online].  
Available: \url{https://www.flywing-tech.com/blog/stm32-vs-esp32-which-microcontroller-is-right-for-your-project/}  
Accessed: July 27, 2025.

\bibitem{Wik2025}
Wikipedia contributors.  
ESP32. In \textit{Wikipedia, The Free Encyclopedia}, June 2025.  
Available: \url{https://en.wikipedia.org/w/index.php?title=ESP32&oldid=1297897793}.  

\bibitem{Patil2024}
Patil, D.; Pournouri, S.  
Evaluating the security of open-source Linux operating systems for unmanned aerial vehicles.  
In: \textit{Cybersecurity Challenges in the Age of AI, Space Communications and Cyborgs},  
Springer Nature Switzerland, Cham, pp. 21--49, 2024.  
\url{https://www.springernature.com/gp/researchers/text-and-data-mining}


\bibitem{b}
ISPRS Archives.  
Aerial surveying UAV based on open-source hardware and software. [Online].  
Available: \url{https://isprs-archives.copernicus.org/articles/XXXVIII-1-C22/155/2011/}  
Accessed: July 27, 2025.

\bibitem{Kim2024}
Kim, Y.; Cho, K.; Kim, S.  
Challenges in dynamic analysis of drone firmware and its solutions.  
\textit{IEEE Access}, vol. 12, pp. 106593--106604, 2024.  
Institute of Electrical and Electronics Engineers (IEEE).  
\url{https://creativecommons.org/licenses/by-nc-nd/4.0/}

\bibitem{Glossner2021}
Glossner, J.; Murphy, S.; Iancu, D.  
An overview of the drone open-source ecosystem.  
arXiv preprint arXiv:2110.02260, Oct. 2021.  
Licensed under CC BY-SA 4.0.  
Available: \url{https://arxiv.org/abs/2110.02260}


\bibitem{Guzman2021}
Guzmán, J. L.; Hägglund, T.  
Tuning rules for feedforward control from measurable disturbances combined with PID control: a review.  
\textit{International Journal of Control}, pp. 1--14, Sep. 2021.  
Informa UK Limited.


\bibitem{Lendek2021}
Lendek, A.; Tan, L.  
Mitigation of derivative kick using time-varying fractional-order PID control.  
\textit{IEEE Access}, vol. 9, pp. 55974--55987, 2021.  
Institute of Electrical and Electronics Engineers (IEEE).  
\url{https://creativecommons.org/licenses/by/4.0/legalcode}

\bibitem{Kumar2016}
Kumar, V.; Rana, K. P. S.  
Some investigations on hybrid fuzzy IPD controllers for proportional and derivative kick suppression.  
\textit{International Journal of Automation and Computing}, vol. 13, no. 5, pp. 516--528, Oct. 2016.  
Springer Science and Business Media LLC.

\bibitem{noauthor_undated}
Nikola Zlatanov.  
Architecture and Operation of a Watchdog Timer. [Online].  
Available: \url{https://www.researchgate.net/profile/Nikola-Zlatanov/publication/295010877_Architecture_and_Operation_of_a_Watchdog_Timer/links/56c65d6b08ae0d3b1b603ee0/Architecture-and-Operation-of-a-Watchdog-Timer.pdf}  
Accessed: July 27, 2025.

\bibitem{Santoso2018}
Santoso, F.; Garratt, M. A.; Anavatti, S. G.; Petersen, I.  
Robust hybrid nonlinear control systems for the dynamics of a quadcopter drone.  
\textit{IEEE Transactions on Systems, Man, and Cybernetics: Systems}, vol. 50, no. 8, pp. 1--13, Aug. 2018.  
Institute of Electrical and Electronics Engineers (IEEE).  
\url{https://ieeexplore.ieee.org/Xplorehelp/downloads/license-information/IEEE.html}

\bibitem{Ratti2011}
Ratti, J.; Moon, J.-H.; Vachtsevanos, G.  
Towards low-power, low-profile avionics architecture and control for Micro Aerial Vehicles.  
In \textit{2011 IEEE Aerospace Conference}, Big Sky, USA, Mar. 2011.  
IEEE.

\bibitem{Katiyar2024}
Katiyar, R.; Chakraborty, P.; Surana, R.; Holla, R.; Sanjana, S.; Acharya, S.; Singh, S.; Agrawal, Y.  
Real-time video PLC using in-painting.  
In \textit{2024 16th International Conference on COMmunication Systems \& NETworkS (COMSNETS)}, Bengaluru, India, Jan. 2024.  
IEEE.


\end{thebibliography}
\end{document}